# ActiveCrowd: A Framework for Optimized Multi-Task Allocation in Mobile Crowdsensing Systems

Bin Guo, *Senior Member*, Yan Liu, Wenle Wu, Zhiwen Yu, *Senior Member*, Qi Han, *Member*

*Abstract*—Worker selection is a key issue in Mobile Crowd Sensing (MCS). While previous worker selection approaches mainly focus on selecting a proper subset of workers for a single MCS task, multi-task-oriented worker selection is essential and useful for the efficiency of large-scale MCS platforms. This paper proposes ActiveCrowd, a worker selection framework for multi-task MCS environments. We study the problem of multi-task worker selection under two situations: worker selection based on workers' intentional movement for time-sensitive tasks and unintentional movement for delay-tolerant tasks. For time-sensitive tasks, workers are required to move to the task venue intentionally and the goal is to minimize the total distance moved. For delay-tolerant tasks, we select workers whose route is predicted to pass by the task venues and the goal is to minimize the total number of workers. Two Greedy-enhanced Genetic Algorithms are proposed to solve them. Experiments verify that the proposed algorithms outperform baseline methods under different experiment settings (scale of task sets, available workers, varied task distributions, etc.).

*Index Terms*—Mobile crowd sensing, platform, task allocation, multiple tasks, ubiquitous computing.

## I. Introduction

The proliferation of smartphones contributes to the prosperity of a novel sensing paradigm called mobile crowd sensing (MCS) [1, 2]. Different from traditional sensing techniques (e.g., wireless sensor networks), MCS utilizes citizens' off-the-shelf smartphones to capture social and urban dynamics. By leveraging human power in-the-loop of sensing and computing process, MCS has become an emerging type of human-machine systems. Numerous novel MCS applications have been studied, such as crowd counting [3], indoor positioning [4], and air pollution monitoring [5]. Commercial MCS sites and applications also appear recently, such as field agent [6], Gigwalk [7], Waze [8], and Millionagents [9].

In MCS applications, we recruit participants or workers to complete them. Considering that the worker candidate pool can be rather large, we should allocate a task to selected and proper workers. Task allocation or worker selection thus becomes an important research issue in MCS.

There have been recently numerous studies on MCS worker selection [10-14]. In these works, they first characterize the needs or constraints of a task (e.g., spatio-temporal contexts, the budget cost, quality of sensing needs), and then select an optimal set of workers to fulfill these needs. However, existing works are mostly single-task oriented, and they do not consider the task allocation problem in a large-scale MCS platform (e.g., at the city-scale) where there can be multiple concurrent heterogeneous tasks requested by different users or systems. Hence, we are faced with new challenges in platform-oriented task allocation in which additional system-level parameters should be considered. This paper aims to address these new challenges by presenting a framework for task allocation in multi-task MCS environments.

MCS tasks are mostly location-dependent, i.e., selected workers need to go to the pre-specified places to complete the tasks. The traveling distance becomes the primary cost for a task worker. Considering task venue correlations, it is more resource efficient if we ask a worker to complete several tasks during her traveling routes. Further, people are willing to undertake several tasks to gain more rewards, assuming that completion of each task is associated with a reward. From the system's perspective, a set of concurrent tasks should be allocated to a selected subset of workers, where each worker can have multiple tasks assigned at a time. This, however, has rarely been investigated in existing studies. Two scenarios of multi-task worker selection are presented below to illustrate the practical values of studying this problem, as shown in Figure 1.

Considering that there is a city-scale MCS platform. Numerous tasks can be published by different requesters during the same time span. Among these tasks, some are required to be completed in a relatively short time period and we call them "time-sensitive tasks". For example, when a rainstorm arrives, many relevant tasks can be published, such as collecting the information of flooded streets, monitoring traffic dynamics, reporting car accidents on the roads. These tasks should be completed as soon as possible due to their emergency. To achieve this, we need to select a subset of workers to complete them intentionally (i.e., by specifically going to the designated task venues, as shown in Fig. 1(a)). In contrast, some sensing tasks are about relatively stable objects and can be completed within a rather long time period (e.g., one or several days), and we call them "delay-tolerant tasks", such as taking photos for the plants of an area to understand the relationship between plant growth and climate change [15], collecting the recent sale information of a shopping mall. For such tasks, we can choose the workers who may visit the task venues during their daily routines, and there is no need to ask the workers to visit those task venues on purpose. An example is shown in Fig. 1(b).

Motivated by these scenarios, we propose ActiveCrowd, which is a framework for multi-task allocation in mobile crowd sensing. We have made the following contributions:

(1) We have proposed the ActiveCrowd framework and studied two common situations for multi-task allocation: worker selection based on workers' intentional movement (i.e., by changing their routines) for time-sensitive tasks (i.e.,

Manuscript received xx; accepted xx. This work was partially supported by the National Basic Research Program of China 973 (No. 2015CB352400), the National Natural Science Foundation of China (No. 61332005, 61373119), the Fundamental Research Funds for the Central Universities (3102015ZY095).

B. Guo, W. Wu, Y. Liu, and Z. Yu are with Northwestern Polytechnical University, Xi'an, China (e-mail: guob@nwpu.edu.cn).

Q. Han is with Colorado School of Mines, US (e-mail: qhan@mines.edu).

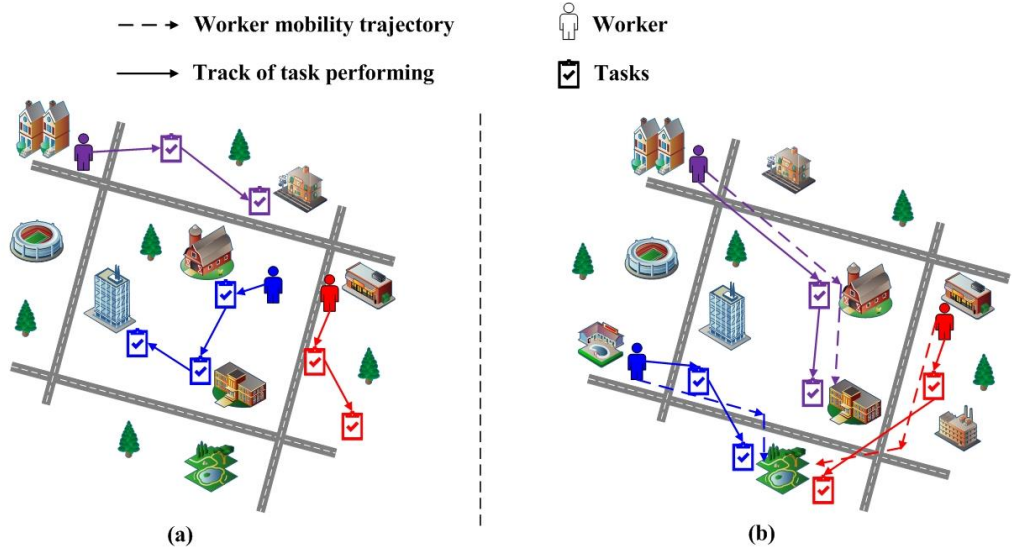

Fig. 1. Two scenarios of multi-task worker selection.

important tasks that should be completed within a short time period), and unintentional movement for delay-tolerant tasks (i.e., completing tasks along the routes of their daily activities).

(2) We have formulated the two multi-task allocation problems. Two greedy-enhanced genetic algorithms, namely GGA-I and GGA-U, are proposed for optimal task allocation.

(3) Experiments over D4D [16], a real-world mobile phone usage dataset with 50,000 users, indicate that our algorithms outperform baselines on both effectiveness and efficiency.

## II. RELATED WORK

With the development of human-in-the-loop systems [17], researchers are now devoted to building generic platforms that can leverage human power for complex problem solving. Mturk [18] is a well-known platform for crowdsourcing, and it has successfully attracted the participation of millions of task requesters and workers. MCS is different from online crowdsourcing in that MCS employs workers to complete tasks in the real world using their sensor-enhanced smart devices. Along with the success of MCS apps such as Waze [8] and SeeClickFix [19], researchers are now interested in developing general-purpose MCS platforms that allow for the publishing and management of tasks with different sensing purposes. Ohmage [20], PRISM [21], and Hive [22] are promising startups towards this direction. These platforms play a major role on task publishing and data collection, but optimized multi-task allocation has not received attention. In these platforms, workers themselves decide which tasks to complete, instead of being matched and selected.

In recent years, there have been studies on optimal worker selection. In the pioneering work, Krause et al.[23] proposed a context-sensitive worker selection approach for optimal community sensing, where both the uncertainty of sensor availability and data privacy are considered when building the selection model. Lappas et al. [10] studied the team-formation problem in consideration of social relations. They form a team of experts to complete a given task aiming to reduce the communication cost among team members. Reddy et al. [11] considered geographic and temporal factors as well as workers' behaviors to select appropriate participants for data collection.

Zhang et al. [12] proposed a participant selection framework to enable requesters to identify participants, aiming to minimize the incentive payments by selecting a small number of participants. Xiong et al. [24] defined a new spatio-temporal coverage metric called $k$-depth, which considers both the ratio of area coverage and the number of contributions in each area. The iCrowd task allocation framework with dual optimal goals (near-maximal $k$-depth coverage and near-minimal incentive payment) were proposed and evaluated. Leveraging the spatio-temporal correlation among the data sensed in different sub-areas, Wang et al. [25] proposed a task allocation framework called CCS-TA, with the aim of minimizing sub-area selection for task allocation. Several state-of-the-art techniques were used for building the framework, including compressive sensing, Bayesian inference, and active learning. However, these works study the worker selection problem for a single task. The optimization goals under multi-task environments are quite different and the existing methods cannot meet the new needs.

Very few works are about multi-task worker selection. Song et al. [13] studied how to select a minimum subset of participants to fully satisfy the quality-of-information (QoI) requirements of concurrent tasks being serviced with total budget constraints. In [14], He et al. studied the problem of allocating location-dependent tasks to appropriate participants to maximize the rewards of the platform. Li et al. [26] proposed dynamic participant recruitment algorithms for heterogeneous sensing tasks, with the aim of minimizing the sensing cost while maintaining certain level of probabilistic coverage. However, they did not take into account the different real-time needs of tasks (time-sensitive vs. delay-tolerant) and the intentional or unintentional movement of workers as presented in our work.

In our work, we propose a worker selection framework for multi-task-oriented MCS platforms. Unlike other studies, we allocate location-based tasks to appropriate participants by considering temporal factors and the spatial movement constraints of workers. In particular, we study the problem for worker selection under two situations: worker selection based on workers' intentional movement for time-sensitive tasks and unintentional movement for delay-tolerant tasks. We mathematically formulate these two problems and further present two greedy-enhanced GA algorithms to address them.

## III. PROBLEM ANALYSIS AND SYSTEM OVERVIEW

In this section, we first analyze the multi-task worker selection problem under two distinct situations, and then describe the proposed ActiveCrowd framework.

### A. Problem Analysis

There are a variety of tasks in MCS, and one of the major differences among them is the time constraints. As our system is more about soft real time tasks, we do not explicitly consider specific deadlines. Instead, based on the urgency of the tasks, we categorize them into two types, namely time-sensitive tasks and delay-tolerant tasks, as presented in the introduction. Time-sensitive tasks need workers to change their original routes and move to the task location specifically. To fully utilize worker resource, each selected worker can complete more than one task. We assume that the incentive rewards to workers are proportional to workers' traveling distance. To minimize the incentive cost, the objective of worker selection based on workers' intentional movement for time-sensitive tasks is to minimize the total movement distance of the selected workers to complete the tasks. In contrast, for delay-tolerant tasks, workers can complete tasks during their daily routines without the need to intentionally change their typical routes. Therefore, we can learn the mobility patterns of workers from their historical traces and select appropriate workers that match the tasks. Different from time-sensitive tasks that need to be completed in a timely manner, delay-tolerant tasks are better completed by a minimum number of workers for better use of worker resource.

### B. The ActiveCrowd Framework

Figure 2 shows our proposed multi-task-oriented worker selection framework. Tasks from task requestors are first classified into time-sensitive tasks and delay-tolerant tasks. Adjacent time-sensitive tasks and delay-tolerant tasks are clustered into groups respectively based on their publishing time and venue. Two greedy-enhanced genetic algorithms are proposed for worker selection to time-sensitive and delay-tolerant tasks, respectively. Details are explained later.

## IV. WORKER SELECTION FOR TIME SENSITIVE TASKS

We focus on the Worker Selection problem for Time-Sensitive tasks (WSTS) in this section.

### A. Problem Formulation

Assuming that the task set to be assigned by the platform is $T=\{t_1, t_2, \cdots, t_n\}$. For the $i$-th ($1 \le i \le n$) task, its location is denoted as $tl_i$ and the number of workers needed is $p_i$. The whole set of workers is $W=\{w_1, w_2, \cdots, w_m\}$. Each worker $w_j$ is located at $ul_j$ ($1 \le j \le m$). To enhance the quality of sensing of tasks, we set a value $q$ on the maximum number of tasks that a worker can be assigned at one time. Due to the urgency of the tasks, we ask the selected workers to specifically go to the designated places to complete the tasks. We simply assume the travel time is proportional to the movement distance. WSTS is to find a scheme that assigns these tasks to a subset of workers in $W$, so that the total distance moved to complete all tasks is minimized.

We need to use workers' current locations and task venue locations to guide the worker selection. Since workers need to travel through city blocks to complete the tasks, we use the Manhattan distance [27] to measure the distance from one task venue to another. The location of tasks and workers is composed of latitude and longitude, and the distance between two locations $l_1$ ($lat_1$, $lon_1$) and $l_2$ ($lat_2$, $lon_2$) can be computed using:

$$dist(l_1, l_2) = |lat_1 - lat_2| + |lon_1 - lon_2| \qquad (1)$$

We use $tu_1, tu_2, \cdots, tu_m$ to denote the tasks assigned to each worker, and $dist(ul_i, tu_i)$ to represent the distance that the $i$-th worker moves to complete task set $tu_i$. Based on these definitions, the problem can be formulated as Eqs. (2)-(4):

$$\min \sum_{j=1}^{m} dist(ul_j, tu_j) \qquad (2)$$

Subject to:

$$|tu_j| \le q (1 \le j \le m) \qquad (3)$$

$$\forall t_i, |\{u_j \mid u_j \in W, t_i \in tu_j\}| = p_i \qquad (4)$$

This is a combinatorial optimization problem. We can prove that it is NP-complete via a reduction from the Travelling Salesman Problem (due to space limit, we do not present it here). The solution space of WSTS is quite large. Supposing that there are $n$ tasks and $m$ workers on the platform, and $p$ workers are required to perform one task. There will be $C_m^p$ worker sets that can be assigned to a task, and $(C_m^p)^n$ assignment schemas for the total of $n$ tasks. The value can be quite large with the increasing of $m$, $q$, and $n$, and thus we use heuristic solutions to solve this problem, as presented below.

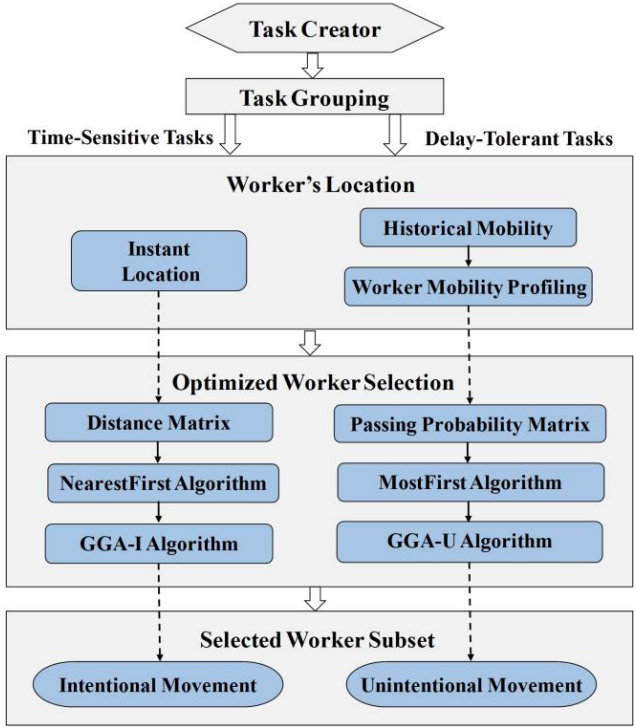

Fig. 2. The Framework of ActiveCrowd.

## B. The NearestFirst algorithm

We first design a greedy algorithm called *NearestFirst* as the baseline. In each iteration, for every tuple <task, worker>, we compute the Manhattan distance between the task and the worker. We find the tuple <*t*, *w*> with the least distance from all the tuples and then assign the task *t* to the worker *w* in this tuple. After that, for task *t*, we decrease its number of persons needed by one. If the number of persons needed becomes zero, we eliminate the task *t* from the task set. For worker *w*, we increase the number of assigned tasks for her by one. If *w*'s number of tasks reaches the limit *q*, we eliminate *w* from the worker pool. The algorithm terminates when either 1) all the tasks have been assigned or no workers can accept a task, or 2) the rest of tasks cannot be assigned to any workers left. The pseudo code is presented in Algorithm 1.

---

**Algorithm 1** The *NearestFirst* algorithm

**Input**: Task set *T*, worker pool *W*, the limit *q* on the number of tasks each worker can obtain.

**output**: The <task, worker> tuple

1: Computing the distance matrix between all tasks and workers according to Eq. (1)
2: While $T \neq \varnothing$ & $W \neq \varnothing$ & $t \in T$ hasn't been assigned to $w \in W$
3: Select the tuple <*t*, *w*> with minimal distance in all tuples and assign the task to the worker
4: Eliminate tasks that have enough workers and workers that have *q* tasks, according to the constraints shown in Eq. (3) and (4)
5: Update distance matrix

---

## C. The GGA-I Algorithm

The *NearestFirst* algorithm is usually sub-optimal because it selects the local best at each step. More elaborated algorithms should be designed. As presented in the problem formulation, the solution space is too large for traditional combinatorial optimization algorithms. Therefore, we turn to Genetic Algorithm (GA) [28], which has been proved to be an effective algorithm for combinatorial optimization. It achieves superior results in a relatively low computation cost. By mimicking the evolution process in the natural world, GA can decrease the influence of the sub-optimal issue by introducing crossover and mutation operators.

In GA, through several generations of selection, crossover and mutation, the initial population converges to the optimal solution or near-optimal solution. Since the solution space of our problem is huge (e.g., hundreds of concurrent tasks and thousands of available workers), individuals in the initial population are very unlikely to be close to the optimal solution, and it is difficult for a traditional GA to get satisfying solutions. Applying some heuristic methods to the basic components of GA has been proved an effective way to remedy this issue [29]. Inspired by this, we propose an algorithm called GGA-I (Greedy-enhanced Genetic Algorithm for Intentional movement) that integrates the heuristic greedy algorithm with GA. GGA-I uses the result of the greedy algorithm as the input for population initiation, the initial population is thus much closer to the optimal solution. As a result, the output of GGA-I is at least as good as *NearestFirst*.

The workflow of GGA-I is illustrated in Fig. 3. In the first step, MCS tasks and candidate workers are given as inputs to the *NearestFirst* algorithm. The result of the greedy algorithm is then used to initialize the GA population. Based on the *NearestFirst* algorithm, we generate the initial population set. Afterwards, the selection, crossover and mutation operations, particularly designed for WSTS, are applied in the evolution process, and we take about a predefined number of iterations (the threshold in Fig. 3) to improve the worker selection result.

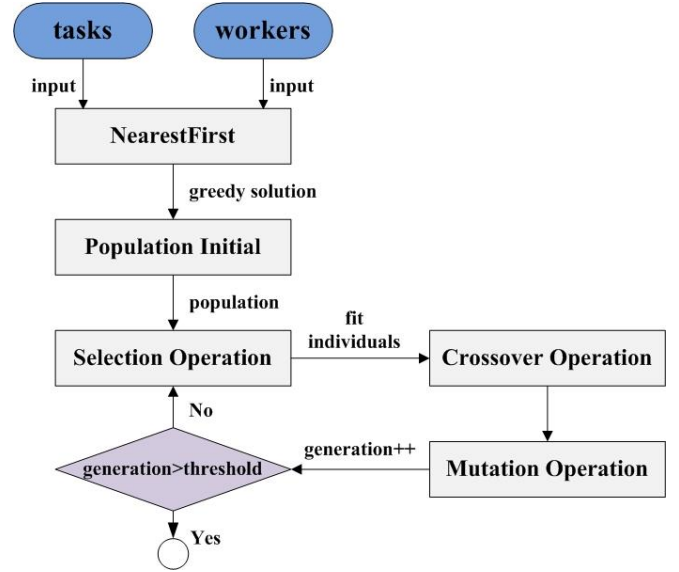

Fig. 3. The workflow of GGA-I.

*(a) Solution representation*. The worker selection problem is the assignment of all the tasks to a subset of workers, and we can adopt the matrix structure to represent a solution. For example, the matrixes in Fig. 4 show the solutions of a case to WSTS with 6 tasks and 8 workers. The rows and columns in the matrix represent workers and tasks, respectively. Each element in the matrix can only be '0' or '1'. If an element is '1', it indicates that the task with the column index of this element is assigned to the worker with the row index of this element. Based on the formulation of WSTS, particularly according to Eqs. (3) and (4), we can define two constraints on the solution representation. We call them task feasibility and worker feasibility:

- *Task feasibility*: the sum of each column equals to the number of people needed by the task with the corresponding column index, i.e., all tasks are with enough workers;
- *Worker feasibility*: the sum of each row does not exceed the limit *q*.

With the two constraints, we say that a solution is feasible when both *task feasibility* and *worker feasibility* are fulfilled.

*(b) Population initialization*. The initial population has significant influence on the result of GA because it is the basis on which the solution is derived and evolved. Usually individuals in the initial population should be uniformly distributed in the solution space. However, the solution space of WSTS is so huge that the random initial population is far from the optimal solution. Given that the result of the NearestFirst algorithm is much closer to the optimal solution than a random solution, we generate the initial population using the result of the NearestFirst algorithm.

*(c) Fitness*. The objective to be minimized is the total distance moved to complete all the tasks. The total distance is the sum of distance moved by each worker. The less the total distance is, the better the fitness is. Each individual stands for one assignment scheme for tasks. As a result, an individual's fitness is related to the total distance that deduced by itself.

Instead of fitness, for WSTS, we use unfitness that refers to the probability one individual is weeded out. An individual's unfitness equals to her total travelling distance. Assuming that in population $I=\{i_1, i_2, \cdots, i_n\}$, the total distance of individual $i_k (1 \le k \le n)$ is denoted as $Td(i_k)$, the unfitness of individual $i_k$ can be calculated as *Eq.* (5):

$$Uf(i_k) = \frac{Td(i_k)}{\sum_{j=1}^{n} Td(i_j)} \qquad (5)$$

The primary role of the unfitness function is to evaluate the fitness of individuals, which can identify the superior and inferior individuals from the population. Note that the superior individuals with lower unfitness than others have higher possibilities to form a new generation, which can inherit excellent genes from the parents. After the evolution process, there is a growing trend that the offspring is better than parents, which means that the final result of *GGA-I* is better than the original result of *NearestFirst*.

*(d) Selection*. The selection operator is to pass individuals with better fitness to the next generation. In other words, individuals that have large unfitness tend to be eliminated. However, it is possible that these individuals also contain some good genes. Hence, we use the roulette wheel [30] to select individuals. The selection operator is as follows: for each generation of population, accumulate all individuals' unfitness; calculate the ratio of each individual's unfitness over the sum, and the result is the probability that the individual should be weeded out.

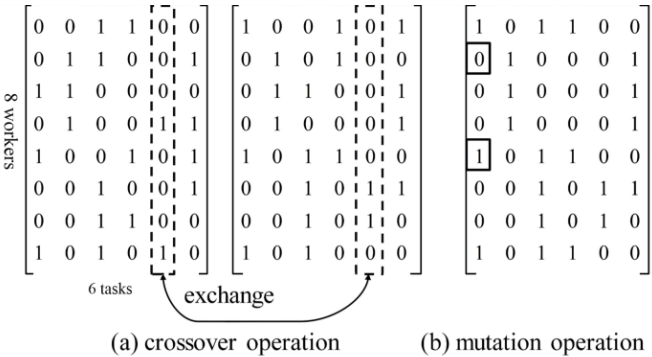

(a) crossover operation     (b) mutation operation

Fig. 4. An illustration of the crossover and mutation operations of GGA-I.

*(e) Crossover*. The crossover operator is responsible for individual reproduction. Child individuals are created through the crossover operation by integrating the parent individuals' genes. For WSTS, we use column exchange for the crossover operator. Because we exchange the columns on the same column index of two parents, the two child individuals created are certainly task feasible. However, the rows to be changed should be carefully selected to ensure that the newly-created children are also worker feasible. Figure 4(a) gives an example of the crossover operation. We separately choose a column on the same column index of two parents at random, and exchange the two columns to generate two children. If the choice does not meet the worker feasibility constraint, we will reselect the columns until *i)* the newly created children are also worker feasible, or *ii)* the selection process attains a given number of times.

*(f) Mutation*. Without the mutation operator, the population may fall into local optimum in the search space. Mutation operator brings the opportunity for individuals to hop to the optimal. For WSTS, we randomly change an element with value '1' to '0', as shown in Fig. 4(b). To ensure that the solution is task feasible, another element in the same column with value '0' should be changed to '1'. After the mutation, the individual should also be validated of worker feasibility.

The pseudo code of GGA-I is shown in Algorithm 2.

| **Algorithm 2** The GGA-I algorithm |
| --- |
| **Input**: The output of *NearestFirst* |
| **output**: The <task, worker> tuple |
| 1: Construct the task assignment matrix $A_{m \times n}$ based on the solution of *NearestFirst*. The matrix conforms to the solution representation. |
| 2: Initialize the population and construct the population matrix $P_{c \times m \times n}$, where $c$ is the number of individuals; |
| 3: Set the iteration count to 0; |
| 4: Computing the unfitness of all individuals in $P_{c \times m \times n}$ according to Eq. (5); |
| 5: Update the currently best individual; |
| 6: Select individuals using the selection operator; |
| 7: Reproduce individuals to fulfill $P_{c \times m \times n}$; |
| 8: Select individuals in $P_{c \times m \times n}$ that are mutated. |
| 9: Increment iteration count by 1, if the iteration count achieves the threshold, stop the iteration; Otherwise go back to step 4; |
| 10: Output the <task, worker> tuple based the best individual. |

## V. Worker Selection for Delay-Tolerant Tasks

The Worker Selection problem for Delay-Tolerant tasks (WSDT) aims to lower the burden of workers by allocating tasks that are on the way of workers. It is achieved through the following two phases:

- *Worker mobility profiling*. In order to predict whether a worker will cover the location point of a task during her movement, we should profile the mobility patterns of workers.
- *Optimal worker selection*. Based on both the locations of existing tasks and the predicted mobility of workers, we select a set of workers to cover them.

### A. Worker Mobility Profiling

Assuming that worker $w$ has historical location records $lr=\{r_1, r_2, \cdots, r_s\}$. Each location record $r_i$ consists of date $rt_i$ and location $rl_i$. The location of task $t$ is denoted as $tl$. Here we suppose that delay-tolerant tasks should be accomplished in 24 hours. Although there are numerous mobility prediction methods [31], we just use the statistical result of a worker's location records to measure the probability that she will pass by the location of a task. It is because that people's daily mobility patterns demonstrate enough regularity for worker selection. The probability that worker $w$ will pass by $tl$ can thus be computed as Eq. (6):

$$p(w,t) = \frac{\left| \{rt_i \mid tl = rl_i\} \right|}{\left| \{rt \mid rt \in \{rt_1, rt_2, \cdots rt_s\}\} \right|} \quad (6)$$

Only when $p(w, t)$ is greater than the passing threshold value $R_{thld}$, task $t$ can be assigned to the worker $w$.

### B. Problem Formulation

Given a task set $T=\{t_1, t_2, \cdots, t_n\}$. For the $i$-$th$ ($1\le i\le n$) task, its location is denoted as $tl_i$ and the number of workers needed is $p_i$. The universal set of workers is $W=\{w_1, w_2, \cdots, w_m\}$. The worker set selected is $WS=\{w_1, w_2, \cdots, w_p\}$. Tasks that each worker $w_i$ gets is $tu_i$. The problem is formulated as *Eqs.* (7)-(9):

$$\min |WS| \quad (7)$$

Subject to:

$$\forall t_i, \ \left| \{w_j \mid w_j \in WS, t_i \in tu_j\} \right| = p_i \quad (8)$$

$$\forall <t_i, w_j>, t_i \in tu_j, \ p(w_j, t_i) \ge R_{thld} \quad (9)$$

Formula (8) ensures that each task has enough workers to complete it. Formula (9) ensures the probability that each worker passes by the task she exceeds the threshold value $R_{thld}$.

We can prove this problem is NP-complete via a reduction from the classic NP-complete Set Cover Problem. Similar to WSTS, the solution space is also large, so we will introduce the greedy algorithm and the enhanced Genetic Algorithm.

Given any subset $WS$ of $W$, we can define a utility function $f(WS)$ of $WS$ to show how much the selected workers in $WS$ can cover the task set. $f(WS)$ is the union set of tasks that all workers in $WS$ can complete. $f(WS)$ is non-negative, monotone and submodular [32] and the proof is given in the Appendix. Similarly to the min-cost coverage problem [33], our problem picks up a minimum set of workers performing sensing tasks to reduce the incentive cost. In addition, the utility function $f(WS)$ we defined is a non-decreasing submodular set function. According to the theory of submodular function, the greedy algorithm can get a near-optimal solution with a constant error bound [33].

### C. The MostFirst Algorithm

We first present a greedy heuristic solution. Since the objective of WSDT is to select the least workers and each worker can take a certain number of tasks, we pick the worker that can accomplish the most tasks at each iteration. We call it *MostFirst,* the pseudo code of which is listed in Algorithm 3.

| **Algorithm 3** The *MostFirst* algorithm |
| --- |
| **Input**: Task set $T$, worker pool $W$, the threshold $r_{thld}$. |
| **output**: The <task, worker> tuple |
| 1:    For each $t \in T$, $w \in W$, compute $p(w, t)$ by Eq. (6). |
| 2:       For each $w_i \in W$, count the number of tasks $c_i$ that $p(w, t)$ is |
|            not fewer than $r_{thld}$ according to Eq. (9). |
| 3.      Select $w_i$ that the corresponding $c_i$ is the most. |
| 4:      If task $t$ have enough workers, remove it from $T$ |
| 5:    If no tasks need to be assigned, stop the iteration. |

Similar to *NearestFirst*, *MostFirst* is also sub-optimal. However, the result can be used as a basis for other methods.

### D. The GGA-U Algorithm

WSDT is also a combinatorial problem, so we can also design a Genetic Algorithm for it to improve the result of *MostFirst*. The workflow of it is the same to GGA-I, which is shown in Fig. 3. We next present the details of several important procedures of GGA-U.

(a) **Solution representation**. In WSDT, once a worker is selected, she can take all tasks whose locations are most likely on her way of typical daily routine. For every worker in the worker candidate pool, we can use '1' to represent that this worker is selected and '0' to represent that this worker is not selected. So we can use a vector to represent the solution of WSDT. The dimension of the vector is the quantity of all workers that can be selected.

(b) **Population initialization**. We first let all individuals be the same as the result of MostFirst. Then we adjust each individual slightly by changing some elements of the vector.

(c) **Fitness**. The objective to be minimized is the total number of workers selected. The less the total number of workers is, the better the fitness is. So we also use unfitness that refers to the probability one individual is weeded out. An individual's unfitness is proportional to the total number of '1' in the individual. Assuming that in population $I=\{i_1, i_2, \cdots, i_n\}$, the number of element '1' that individual $i_k(1\le k\le n)$ has is denoted as $c_k$, the unfitness of individual $i_k$ can be calculated as Eq. (10):

$$Uf(i_k) = \frac{c_k}{\sum_{j=1}^{n} c_j} \quad (10)$$

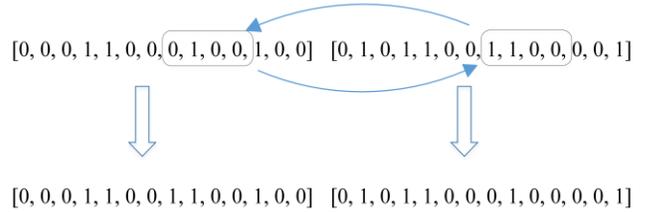

Fig. 5. Illustration of the crossover operation of *GGA-U*

(d) **Selection**. Individuals that have large unfitness tend to be eliminated. The roulette wheel method [30] is used to select individuals in population generation, similar to GGA-I.

(e) **Crossover**. For individual reproduction, we swap pieces of two parent individuals. The process is illustrated in Fig. 5.

(f) **Mutation**. We randomly select elements in individual's representation and change it to the opposite.

## VI. EVALUATION AND DISCUSSION

In this section, we evaluate the performance of our proposed algorithms using a large-scale real-world dataset. The impacts of number of tasks/workers and the task distribution on worker selection, are also studied.

### A. The Dataset

To model the location and movement of workers in a metropolitan area, we utilized the D4D dataset [16]. It contains individual call detail records for over 50,000 customers of Orange Group during two weeks in Ivory Coast. Each record consists of user id, the date/time of the phone call, and the cell tower from where the phone call is made. The total number of cell towers is 1231. Considering that cell towers in the capital Abidjan are the densest (347 in 1231), we mainly use the records in the Abidjan area in the experiments.

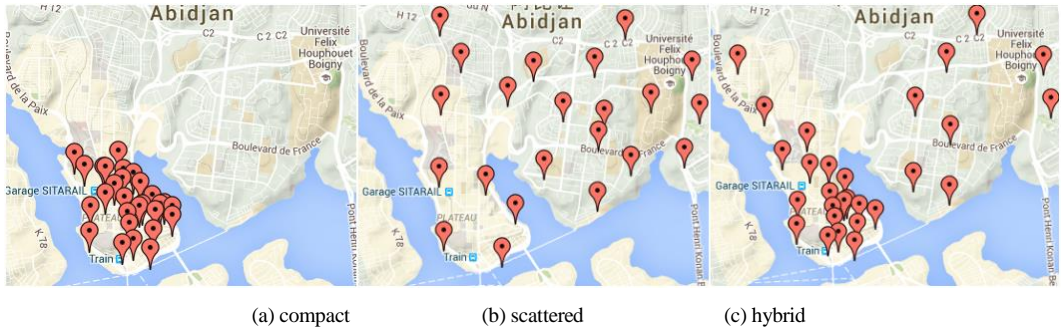

(a) compact       (b) scattered       (c) hybrid

Fig. 6. An illustration of the three types of task distribution

Every cell tower has its id, latitude and longitude, so we can extract the users' mobility traces related to cell towers. The locations of workers/tasks are all related to the location of cell towers, and we use cell tower id to represent the location of tasks/workers. The distance that a worker moves from one cell tower to another is the distance between these two cell towers. For simplicity, we assume that the distance that a worker takes tasks within the same cell tower is zero.

### B. Experiment Settings and Baselines

**(1) Experiment Settings**

We first present the settings of our experiments. For WSTS, to ensure quality of sensing, the number of workers needed by each task is randomly generated between 2~4, and the upper bound $q$ on the number of tasks that each worker can be assigned at a time is set to '3'. To measure the completion time of tasks, we assume that the moving speed of a worker is 70 meters per minute. In each experiment, locations of tasks are randomly distributed to a group of cell towers within a target area, and the workers to be selected are users in D4D who make phone calls around these cell towers within a pre-specified one-hour time window (tasks are also assumed published during that period).

In WSDT, workers are selected based on their mobility patterns. Once the task set is determined, we use the latest 10-day call records to compute the probability that each worker will pass by the task location in the next 24 hours (we assume that each task can be completed within one day). If a worker actually passes by the task location in the next day (in call records she made phone calls under the cell tower that the task is in), we say that she can accomplish that task. The number of workers that each task needs is also set to 2~4.

To measure whether our method performs well under different spatial distribution of tasks, we adopt three types of task distribution.

- **Compact distribution**. The tasks are relatively close to each other on the map. For example, sensing of a local crowd event like a big music festival would lead to a compactly distributed task.
- **Scattered distribution**. In this distribution, the tasks are scattered over the target area. The generation of a crowdsourced air pollution map is a typical task with scattered distribution.
- **Hybrid distribution**. The distribution of tasks is a combination of the above two modalities. There are many tasks with hybrid distribution, e.g., reporting the traffic dynamics of important places in the urban area. The places close to the downtown area are more compact than suburb areas.

The illustration of the three types of distributions (in the Abidjan area) used in our experiments is given in Fig. 6.

**(2) Baselines and Evaluation Metrics**

The following algorithms are used as baselines.

- **GA** - It refers to the traditional Genetic Algorithm, without the usage of heuristic inputs.
- **EA** - To evaluate the performance of GGA-I, we also implement the enumeration algorithm (EA) which can search the whole solution space, i.e., an optimal solution. Given that the complexity of WSTS increases exponentially with the increase of the task and worker number, we only run it at a relatively small scale.
- **GYPSO** - The Particle Swarm Optimization (PSO) [34] is another popular evolutionary algorithm. The start of the evolution is a group of random solutions, the same as GA. The difference of them is that PSO does not have the selection/crossover/mutation operators. In contrast, individuals in PSO update themselves by tracing the global and individual optimal solutions [34]. For comparison, the individual initiation of PSO is also based on *NearestFirst*, and we use "GYPSO" to denote it.

Several metrics are proposed for evaluation.

- For WSTS, the *total movement distance* of selected workers (i.e., the optimization goal) is the chief indicator for comparison. As an optimization problem, the *computation complexity* is also important. Finally, for time-sensitive tasks, we also wish that the *average completion time* of each task is as short as possible.
- For WSDT, the *accuracy of mobility prediction* has great influence on the task completion ratio. The *number of workers selected*, as the optimization goal, is another important indicator. In addition, we investigate the worker selection results under various *task distributions*.

In terms of the similar structure of GGA-I and GGA-U, we only compare GGA-U with the greedy approach – *MostFirst*.

### C. Evaluation of GGA-I

The optimization goal of WSTS is to minimize the total distance for completing all tasks. Fig. 7 shows the results under several different situations. Here, '3t5w' means that the number of tasks is 3 and the candidate worker number is 5. For each situation, we conducted 20 times of experiments to get the average value.

In Fig. 7, we present the performance comparison on total distance of selected workers between GGA-I and baselines. GGA-I outperforms *NearestFirst* and GYPSO. The reason that GGA-I performs better than GYPSO is that GYPSO tends to fall into local optimum (crossover/mutation operators are not used). GGA-I performs much better than GA, proving the usefulness of the integration with heuristic inputs. We can see the result of GGA-I is close to EA, which implies that the performance of GGA-I is close to the optimal solution.

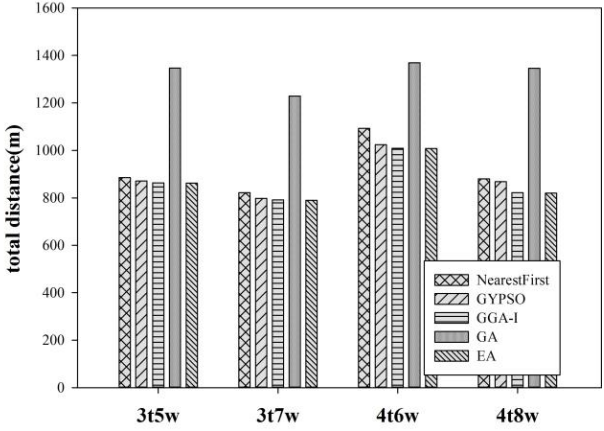

Fig. 7. The total distance moved under different situations.

The computational complexity of the five algorithms under different worker pool size is shown in Fig. 8. The number of tasks to be assigned is set to 4. We can find that though EA can minimize the total distance, its computation complexity grows rapidly with the increase of candidate worker number, and the value is much higher than the other algorithms. GGA-I is a bit more computational intensive than *NearestFirst*, GYPSO and GA, but it achieves better solutions than them. Therefore, GGA-I performs well when both effectiveness and efficiency are taken into consideration.

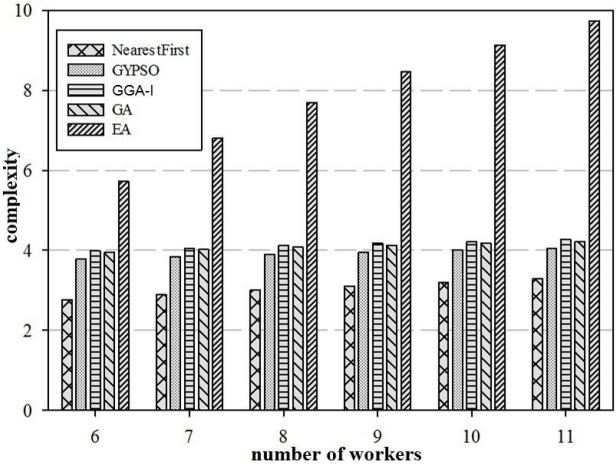

Fig. 8. The logarithmic computation complexity of each algorithm.

The solution space grows exponentially with the increasing number of tasks and workers. However, the computation complexity in GGA-I does not increase dramatically when the number of tasks and workers increases, as shown in Table 1. Therefore, when a large number of tasks are being executed concurrently, GGA-I works efficiently on selecting proper workers.

TABLE I. TIME COMPLEXITY OF GGA-I

| Tasks-workers | 10t20w | 20t40w | 30t60w | 40t80w | 50t100w |
|---|---|---|---|---|---|
| Runtime per generation(s) | 0.034 | 0.043 | 0.056 | 0.072 | 0.089 |

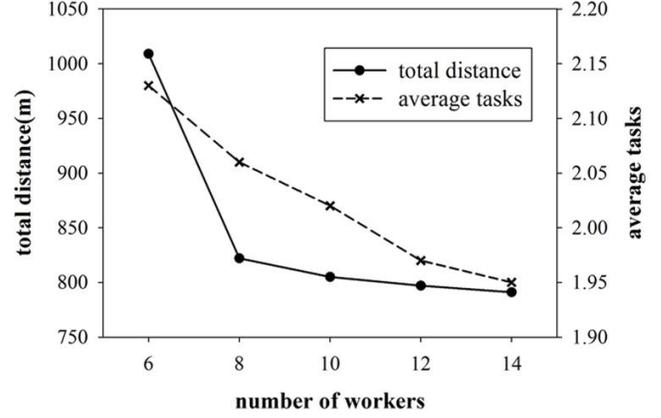

Fig. 9. The total distance and average number of tasks each worker gets with the increase of candidate worker number in GGA-I.

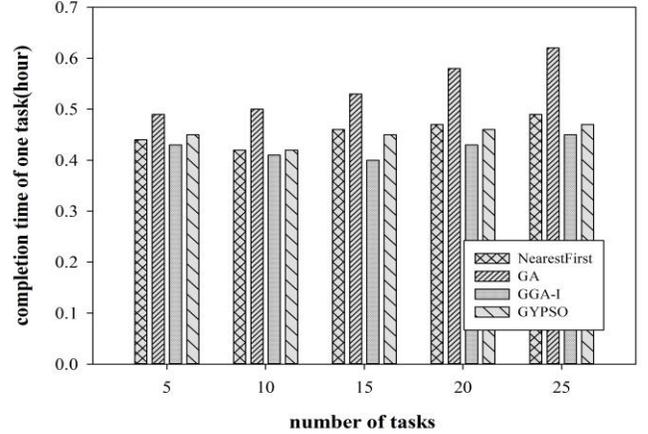

Fig. 10. The average completion time of a task.

For a given number of tasks, with the rising of the candidate worker number, there can be more proper workers. The total distance also becomes smaller because there may be a better choice among the newly added workers. However, the average number of tasks each worker gets becomes less since more workers share the same task set. These findings can be seen from Fig. 9. However, we find that with the increase of candidate workers, the total distance decreases very slowly. It indicates that for a given set of MCS tasks, a reasonable scale of candidate workers is enough. This conclusion can help us better manage worker resource and perform platform-level scheduling.

The completion time is important for time-sensitive tasks, and we expect the completion time is as short as possible in WSTS. In Fig. 10, we compare the average completion time of a task between GGA-I and other algorithms under different task number. We find that GGA-I outperforms *NearestFirst*, GA, and GYPSO on the average completion time, mostly because that the total movement distance in GGA-I is the least.

TABLE II.    NUMBER OF SELECTED WORKERS

(a) compact distribution     (b) scattered distribution     (c) hybrid distribution

| Task Set | GGA-U | *MostFirst* |
|----------|-------|-------------|
| $R_{thld}$ = 90% | | |
| 1 | 36 | 40 |
| 2 | 36 | 42 |
| 3 | 39 | 45 |
| **avg.** | **37** | **42.3** |
| $R_{thld}$ = 80% | | |
| 1 | 32 | 35 |
| 2 | 31 | 34 |
| 3 | 31 | 36 |
| **avg.** | **31.3** | **35** |

| Task Set | GGA-U | *MostFirst* |
|----------|-------|-------------|
| $R_{thld}$ = 90% | | |
| 1 | 38 | 41 |
| 2 | 37 | 44 |
| 3 | 40 | 48 |
| **avg.** | **38.3** | **44.3** |
| $R_{thld}$ = 80% | | |
| 1 | 34 | 36 |
| 2 | 34 | 37 |
| 3 | 32 | 37 |
| **avg.** | **33.3** | **36.7** |

| Task Set | GGA-U | *MostFirst* |
|----------|-------|-------------|
| $R_{thld}$ = 90% | | |
| 1 | 38 | 41 |
| 2 | 36 | 42 |
| 3 | 40 | 47 |
| **avg.** | **38** | **43.3** |
| $R_{thld}$ = 80% | | |
| 1 | 34 | 36 |
| 2 | 33 | 36 |
| 3 | 33 | 35 |
| **avg.** | **33.3** | **35.7** |

## D.  Evaluation of GGA-U

We first validate the effectiveness of the mobility profiling method, and then conduct several experiments under different task distributions and the passing threshold value $R_{thld}$ . The comparison on the number of selected workers between GGA-U and *MostFirst* is also presented.

The accuracy of predicted mobility has impact on the completion ratio of tasks because it will affect the formation of the candidate worker pool. We measure the predicted and practical probability under different thresholds. The predicted probability is the average probability that the selected worker will pass by the task location based on the *prediction* over her historical mobility. The practical probability refers to the average possibility that the selected worker *actually* passes the task location based on the call records in the next day. That's to say, the practical probability is the ratio of the number of tasks completed to the number of tasks assigned. As shown in Fig.11, it is clear that the value of predicted probability is mostly similar to the practical probability, revealing the effectiveness of our method. The predicted/practical probability values are all bigger than $R_{thld}$ because the threshold is the least value of passing probability that a worker can be selected to complete a task.

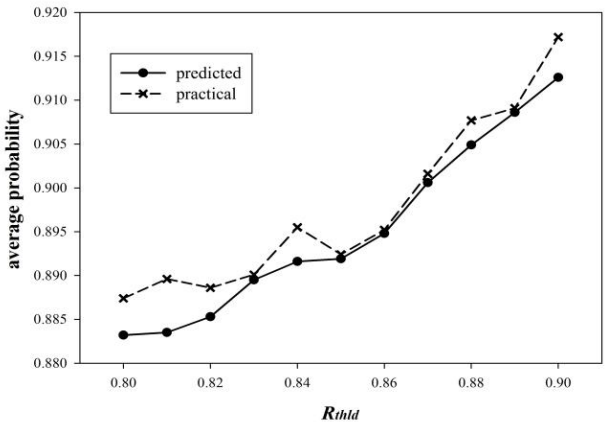

Fig. 11. The predicted probability and practical probability

We study the number of selected workers under three situations: the compact distribution, the scattered distribution, and the hybrid distribution. There are 20 tasks needed to be assigned, and we set two different values to $R_{thld}$. For each situation, we experiment on three different task sets (labeled as *1-3*), which are generated according to the task distributions respectively. The experimental results are shown in Table 2. It is clear that GGA-U outperforms *MostFirst*, and the number of selected workers decreases under a small $R_{thld}$. The reason is that there are more options for workers after lowering the threshold. In addition, from Fig. 11 we can find that increasing the value of $R_{thld}$ does not promote much on the predicted and practical probability. Therefore, using a relative lower $R_{thld}$ can result in a small number of selected workers, without a significant impact to the completion ratio of tasks.

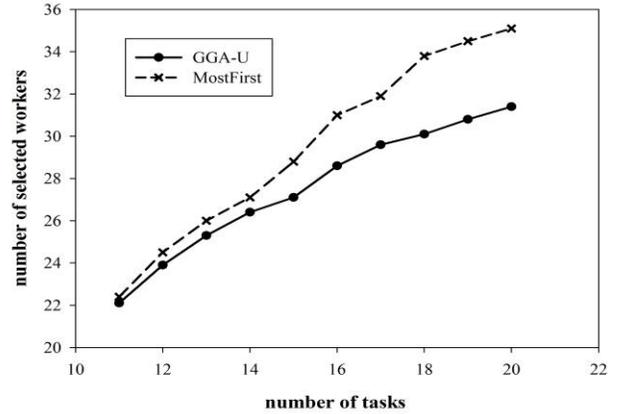

Fig. 12. The number of selected workers with the increase of the number of tasks. ($R_{thld}$ = 80% )

In general, the number of selected workers under the compact distribution is the fewest among the three different distributions. It is mainly because that people tend to shift within a local area in their daily activities rather than a wide-scale area (e.g., the example in Fig. 6(b)). When the tasks are in scattered distribution, they have less probability for co-occurrence in a worker's route. Therefore, more workers should be recruited in such distributions. Further, GGA-U performs well than *MostFirst* under different task distributions, in terms of the number of selected workers. We can also find from Fig. 12 that the number of selected workers increases with the increase of task number, and the increase ratio of *MostFirst* is greater than GGA-U.

## VII.  DISCUSSIONS

This section discusses issues that are not reported or addressed in this work due to space and time constraint, which can be added to our future work.

*Heterogeneous tasks*. There are broadly two types of MCS tasks, namely one-shot tasks and recurrent tasks. Quite many tasks are one-shot tasks (SmartEye [35], SmartPhoto [36]), but there are tasks that need recurrent sampling, especially in the environment monitoring field. There have been recent studies on worker selection for recurrent MCS tasks [12] [26]. However, they do not consider intentional worker selection. We will extend ActiveCrowd to support it in the future.

*Optimization goals.* The current work aims at either minimizing the total distance of workers or minimizing the worker set for the sake of platform running. Other factors are also important to MCS, such as incentives and budget. We will take into account these factors in our future work.

*Methods for worker selection.* In this paper, we present greedy-enhanced GA methods to solve the multi-task worker selection problem. Experiments demonstrate that our methods are effective and efficient. Optimization is a widely-studied research area and we are trying to explore other advanced as well theoretical-founded methods to solve our problems. For example, submodularity [32] has been proved a good feature for combinatorial optimization problems, and we will study its applicability to our problems and may propose new methods.

*Worker profiling.* Quality of sensing is an important problem in MCS. It is largely impacted by the expertise/experience of workers. That's to say, if a worker has more knowledge about the sensing task, it is more possible that the task can be completed efficiently and of high quality. However, when this factor is considered, several challenges may arise: 1) the expertise/experience of workers is not easy to obtain; 2) the optimization problem will be more challenging because more constraints are introduced. Similar to most crowdsourcing systems, we can ask workers to provide their profiles when they register in our system. Furthermore, inspired by the studies on expert finding from social media [37, 38], we can also learn human expertise or experience from user-generated data in social media. For example, user check-ins in location-based social networks allows us to learn the familiarity of a user to the task venue; online posts of a user can expose her interests or expertise. However, social-media-based profile mining is often time-consuming, and we can perform it offline to enrich or validate user-provided profile information. We plan to study this problem in our future work.

*Large-scale user study.* In this work, the evaluation of our algorithms is based on a real-world human mobility dataset. We are now collaborating with the local government to build a city-level MCS platform for citizens to report municipal problems (e.g., road collapse, public facility damage, and noise disturbance), urban dynamics (e.g., traffic jams, accidents, public information [39]), etc. We intend to introduce our multi-task allocation framework in the platform and make large-scale user studies, which will help identify practical issues and improve our framework.

## VIII. CONCLUSION

We have studied the worker selection problem under multi-task MCS environments. Two common situations are identified: intentional-movement-based selection for time-sensitive tasks and unintentional-movement-based selection for delay-tolerant tasks. The optimization goal for the two situations varies and we propose two greedy-enhanced genetic algorithms GGA-I, GGA-U to address them. Experimental results show that, compared to baseline methods such as GA, EA, GYPSO, our algorithms can better meet the optimization needs with high efficiency. Our methods always outperform the basic greedy algorithm and GA under various circumstances. As for future work, we will consider other factors that may affect the selection of workers in multi-task MCS environments. New optimization methods and theoretical foundations will also be studied and leveraged.

## APPENDIX

*Problem:* The utility function $f(WS)$ of the selected worker set WS is non-negative, monotone, and submodular. $f(WS)$ is the union set of tasks that all workers in $WS$ can complete.

*Proof*: Since $f(WS)$ is a set of tasks, $f(WS)$ is obviously non-negative. Assume that $WS_1 \subseteq WS_2$, $f(WS_1) \subseteq f(WS_2)$. So $f(WS)$ is monotone. Now consider $f(WS_2 \cup \{w\}) - f(WS_2)$, which is the increase of tasks completed due to adding $w$ to $WS_2$. If selecting worker $w$ that does not increase any on $f(WS_2) - f(WS_1)$, $f(WS_2 \cup \{w\}) - f(WS_2) = f(WS_1 \cup \{w\}) - f(WS_1)$. On the other hand, if $w$ completes a task that is in $f(WS_2) - f(WS_1)$ while not in $f(WS_1)$, then $f(WS_2 \cup \{w\}) - f(WS_2) < f(WS_1 \cup \{w\}) - f(WS_1)$. Therefore overall $f(WS_2 \cup \{w\}) - f(WS_2) \leqslant f(WS_1 \cup \{w\}) - f(WS_1)$ and $f$ is submodular.

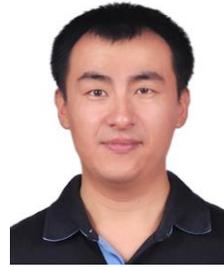

**Bin Guo** is a professor from Northwestern Polytechnic University, China. He received his Ph.D. degree in computer science from Keio University, Japan in 2009 and then was a post-doc researcher at Institut TELECOM SudParis in France. His research interests include ubiquitous computing, mobile crowd sensing, and HCI.

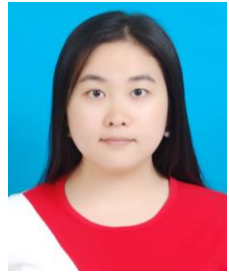

**Yan Liu** is currently a master student in Northwestern Polytechnical University, China. Her research interest is mobile crowd sensing and intelligent systems.

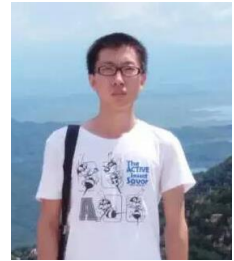

**Wenle Wu** is a master student in Northwestern Polytechnical University, China. His research interest is mobile crowd sensing and ubiquitous computing.

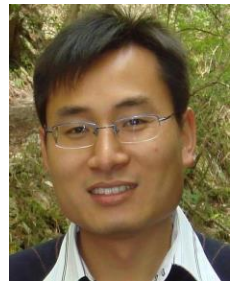

**Zhiwen Yu** is currently a professor and vice-dean at the School of Computer Science, Northwestern Polytechnical University, China. He has worked as an Alexander Von Humboldt Fellow at Mannheim University, Germany from Nov. 2009 to Oct. 2010, a research fellow at Kyoto University, Japan from Feb. 2007 to Jan. 2009. His research interests cover ubiquitous computing and HCI.

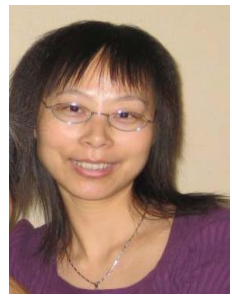

**Qi Han** is an associate professor of computer Science from Colorado School of Mines, US. She obtained her Ph.D. degree in Computer Science from the University of California-Irvine in August 2005. Her research interests include mobile crowd sensing and ubiquitous computing.